# Nontrivial topology of bulk HgSe from the study of cyclotron effective mass, electron mobility and phase shift of Shubnikov-de Haas oscillations

S.B. Bobin, A.T. Lonchakov, V.V. Deryushkin, V.N. Neverov

*M.N. Miheev Institute of Metal Physics of Ural Branch of Russian Academy of Sciences, 620108, Yekaterinburg, Russia*

**Abstract**

In this paper, the authors report the results of an experimental study of effective mass, electron mobility and phase shift of Shubnikov-de Haas oscillations of transverse magnetoresistance in an extended electron concentration region from $8.8 \times 10^{15}$ cm$^{-3}$ to $4.3 \times 10^{18}$ cm$^{-3}$ in single crystals of mercury selenide. The revealed features indicate that Weyl semimetal phase may exist in HgSe at low electron density. The most significant result is the discovery of an abrupt change of Berry phase $\approx \pi$ at electron concentration $\approx 2 \times 10^{18}$ cm$^{-3}$, which we explain in terms of a manifestation of topological Lifshitz transition in HgSe that occurs by tuning Fermi energy via doping.

## I. Introduction

The increasing interest in Weyl semimetals (WSMs) in recent years has not only been in terms of fundamental research but also with respect to their potential for future engineering applications [1,2]. A WSM is a three-dimensional (3D) topological phase protected by a crystal symmetry and characterised by linear dispersion around pairs of nodes having opposite chirality. These so-called Weyl nodes are separated in momentum space; their only connection is through the sample boundary by exotic non-closed surface states known as Fermi arcs [3]. In paper [4] we reported spectacular transport properties in a HgSe sample with low ($2.5 \times 10^{16}$ cm$^{-3}$ at $T=$ 4.2K) electron concentration, including strong non-saturating transverse magnetoresistance (MR), negative longitudinal MR, which could be related to a chiral anomaly, and unusual for gapless semiconductors maxima of the temperature dependence of electrical resistivity. These features indicate the possibility of the coexistence of a WSM phase with a trivial gapless semiconductor phase in this material with zinc-blend structure. There are other experimental (ARPES investigation [5]) and theoretical [6, 7] justifications for proposing mercury selenide, a material with strong spin-orbit coupling, as a possible candidate in non-centrosymmetric WSMs. Time reversal symmetry forces the total amount of a Weyl nodes in non-centrosymmetric system to be a multiple of four, e. g. there must be at least two pairs of Weyl nodes with opposite chirality. On the other hand, cubic $T_d$ symmetry of HgSe includes either mirror planes or 2-fold and 4-fold rotation axes. That is, the high cubic symmetry of HgSe could allow the location of at least four pairs of Weyl nodes with opposite chirality in a single ($k_x$, $k_y$) plane instead of the two pairs in a single plane in strained HgTe [7] with lower tetragonal $D_{2d}$ – symmetry.

Thus, mercury selenide is of significant interest for topological condensed matter physics as it presumably has four Weyl node pairs of the same type in bulk and surface topological Fermi arcs. This material can be considered as a simpler or "ideal" non-centrosymmetric WSM, the study of which is important for elucidating the exotic transport properties that other non-centrosymmetric WSMs with a more complex Fermi surface topology exhibit [1]. The aim of this paper is to provide arguments for the possible existence of the WSM phase in HgSe based on the concentration dependence of effective mass, Hall mobility and the Berry phase, which is a principal topological parameter of WSM kinetics.

### II. Crystal structure and experimental details

Mercury selenide single crystals were grown using the Bridgman technique. Crystal structure of the HgSe ingots was investigated using the high-resolution x-ray diffractometer with Cu anode radiation in the Collaborative Access Center of M.N. Miheev Institute of Metal Physics of Ural Branch of Russian Academy of Sciences. Rietveld refinements of the diffractograms were performed using the software package HighScore Plus 4.1. A typical room temperature XRD patterns of HgSe powder is shown in fig. 1.

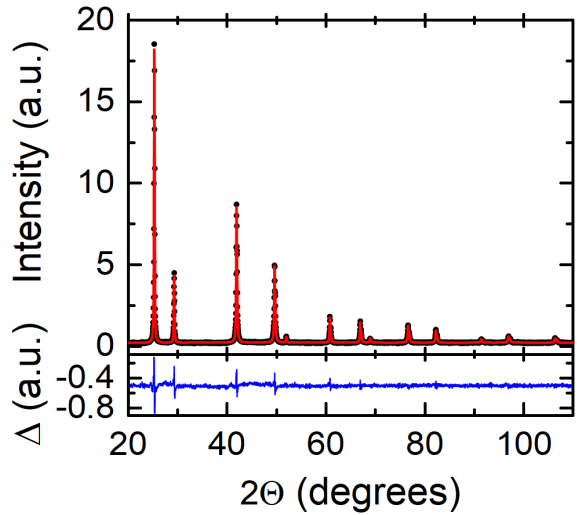

*FIG. 1. Upper panel: room temperature X-ray diffraction patterns of HgSe powder (symbols – experiment, solid line - fitting). Lower panel: difference between the fitting and experiment.*

As a result of the diffractogram analysis, $F\bar{4}3m$ space symmetry group of the zinc-blende structure of HgSe was confirmed with unit cell parameter 6.0875 Å and with Wyckoff positions 4a (000) and 4c (¼,¼,¼) related to Hg and Se, respectively.

Samples under investigation were cut from the homogeneous part of the single crystal ingots perpendicular to the direction of growth. Electron concentration in the range of ($10^{16} \lesssim n_e < 10^{18}$)

cm$^{-3}$ was obtained by varying the annealing time of the samples in Se vapour from ≈ 24 to ≈ 600 hours at temperatures of (150 – 200)°C [8]. Samples with $n_e$ ≈ (1 – 5)×10$^{18}$ cm$^{-3}$ were as-grown. In order to increase $n_e$ above 2×10$^{18}$ cm$^{-3}$, samples were doped with donor impurities Ga or Fe [10]. All samples had the shape of a rectangular parallelepiped whose dimensions varied between ≈ (3.5×1×0.7) mm$^3$ and ≈ (8×2×1) mm$^3$ and which were not specifically oriented. The procedure of etching and electrical contact preparation is described in [4].

Transverse MR $\rho_{xx}$ was measured using a standard four-probe method at temperatures 2, 4.2, 10 and 20K in a magnetic field up to 12T. To eliminate the nonsymmetrical effect of the electrical contacts all measurements were carried out in two opposite directions of the magnetic field, then the $\rho_{xx}(B)$ was obtained by averaging over the magnetic field direction $\rho_{xx}(B) = [\rho_{xx}(+B) + \rho_{xx}(-B)]/2$.

### III. Results and discussion
### A. Electron effective mass and Hall mobility

The effective mass $m_e(n_e)$ of the electrons in HgSe in the range of low $n_e$< 1×10$^{17}$ cm$^{-3}$ attracted our interest due to a lack of experimental data in the literature. This particular region is especially relevant to a study of the topological properties of HgSe [4] since, in condensed matter physics, Weyl fermions are low-energy excitations. The cyclotron effective mass $m_e$ of the electrons was deduced from the ratio of amplitudes of Shubnikov de Haas (SdH) oscillations of transverse MR $\rho_{xx}(B)$ measured at different temperatures. A detailed description of the method of obtaining $m_e$ from the analysis of the SdH oscillations is given in [4]. SdH oscillations were extracted from $\rho_{xx}(B)$ after subtracting a smooth MR background. The oscillatory component of transverse MR for several HgSe samples is shown in Fig. 2. The insets demonstrate the fitting result of the ratio of oscillation amplitudes versus $1/B$ based on Lifshitz-Kosevich (LK) formula which yields the cyclotron effective mass [4]. In order to determine the electron concentration $n_e$, we used the formula:

$$n_e = \frac{(2eF)^{3/2}}{3\pi^2 \hbar}, \qquad (1)$$

where $F$ – FFT frequency of SdH oscillations, $e$ is the absolute value of the electron charge, $\hbar$ is the Plank constant ($\hbar = h/2\pi$).

Fig. 3 demonstrates our findings for the electron mass ratio $m_e/m_0$ as a function of $n_e$ as well as the data for $m_e/m_0$ from [11,12], where $m_0$ is a free electron mass. It can be seen that, within the margin of error, effective mass does not depend on electron concentration in the range of $n_e \lesssim 1\times10^{17}$ cm$^{-3}$ down to the lowest concentration for our samples ≈ 9×10$^{15}$ cm$^{-3}$. On the one hand, this important result is not consistent with the generally accepted view of the non-parabolicity of the isotropic band $\Gamma_8$ in HgSe [13]. On the other hand, it can be interpreted as a confirmation of hypothesis [4] for the predominance of a topological semimetal phase in HgSe kinetics at low $n_e$ because a rather long annealing time could lead to a depopulation of the conductivity band $\Gamma_8$. It can be assumed that the Weyl fermion dispersion relation $\varepsilon(k)$ is not an ideal Dirac-like dispersion, which generally deviates from the ideal Dirac cone, but can be described as in [14,15]:

$$\varepsilon(k) = \varepsilon_W + v_F \hbar k + \frac{\hbar^2 k^2}{2m_e}, \qquad (2)$$

where $\varepsilon_W$ is the energy at the Dirac point $k_W$, and $v_F$ is the velocity at the Dirac point. As a result of (2), we detect nondependent on $k$ electron effective mass at low $n_e$. Only at $n_e \gtrsim 5\times10^{17}$ cm$^{-3}$ can the dependence $m_e(n_e)$ be satisfactorily described using Kane's [16] non-parabolic isotropic dispersion law (solid line in Fig. 3). To fit the dependence $m_e(n_e)$ in two-band approximation (see formula (15) in [11]), we used the following parameters: $\varepsilon_g = \varepsilon(\Gamma_6) - \varepsilon(\Gamma_8) = -0.23$eV, which is the energy gap between light hole band $\Gamma_6$ and conduction band $\Gamma_8$ for trivial HgSe band structure, and $P = 7.7\times10^{-8}$ eV cm which is the interband momentum matrix element. Thus, we can assume that, at a sufficiently large $n_e$, the concentration dependence of the effective mass points to the dominance of a trivial gapless semiconductor phase (conduction band $\Gamma_8$).

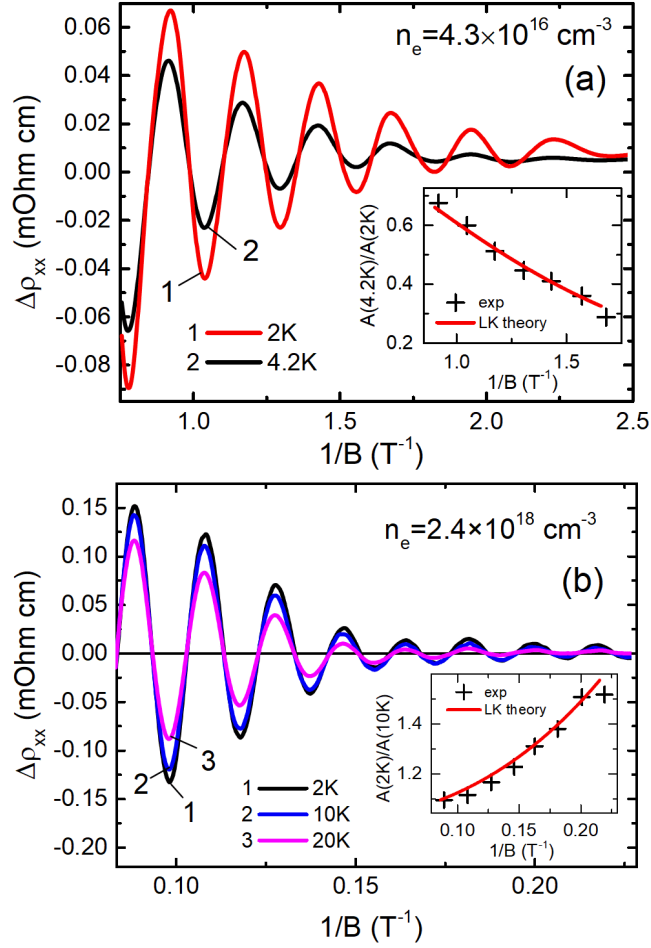

FIG. 2. *The oscillatory component $\Delta\rho_{xx}$ of the transverse MR as a function of $1/B$ after substracting the background of $\rho_{xx}(B)$ at different temperatures for the HgSe sample with $n_e = 4.3\times10^{16}$ cm$^{-3}$ (a) and $2.4\times10^{18}$ cm$^{-3}$ (b). The insets show the dependence of the oscillation amplitude ratio versus $1/B$ for the studied temperatures (symbols). The solid line is the fit to the LK formula used to determine the effective mass [4].*

Unlike the effective mass, which is a parameter of the band spectrum, the mobility $\mu = e\tau_{tr}/m_e$ additionally characterises the electron scattering processes through the transport relaxation time $\tau_{tr}$. We experimentally determined electron mobility using formula for Hall mobility $\mu_H = \sigma_0/n_e e$, where $\sigma_0$ is zero-field conductivity.

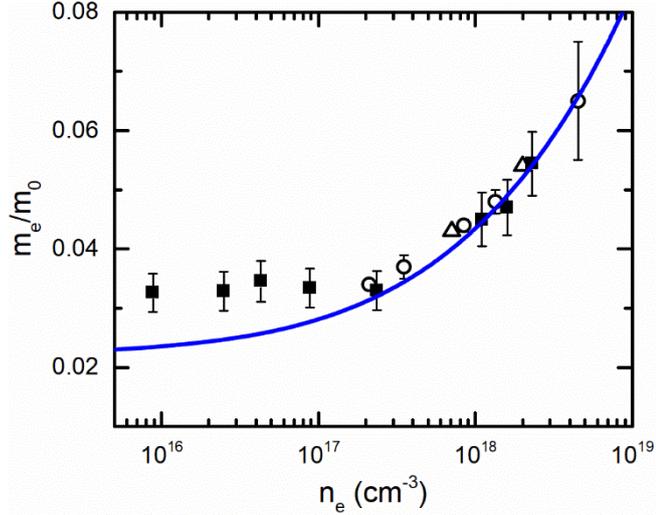

FIG. 3. *Cyclotron effective mass in HgSe normalised to $m_0$ as a function of electron concentration at 4.2K. Filled squares – this work, open circles are from [11], open triangles are from [12], the solid line is the calculated effective mass for non-parabolic isotropic dispersion in two-band approximation.*

Fig. 4 shows the concentration dependence of $\mu_H$ at $T$ = 4.2K, consisting of the data of [9, 11, 17-19] combined with our own findings. After analysing this data, we distinguished three regions of $\mu_H(n_e)$ (Fig. 4). Region I corresponds to $n_e > n_{LT} \sim 10^{18}$ cm$^{-3}$, where $n_{LT}$ is the concentration of the Lifshitz transition (LT), the discovery of which in HgSe is discussed below. Electron topological LT [20] is the abrupt topological change in the connectivity of the Fermi surface of material, giving rise to anomalies of its physical properties. With regard to the WSMs, bulk LT should be considered as a recombination of two non-trivial Fermi surfaces enclosing one Weyl node into a single trivial Fermi surface that encloses both Weyl nodes of opposite chirality [21-23]. This means that chirality is not well-defined, i.e. the total Berry flux through the Fermi surface equals zero. In Region I, the dependence $\mu_H(n_e)$ seem to be reasonably well described by the theory of ionised donor scattering (solid line on Fig. 4 from [18]) and transport properties should be determined by the trivial conduction band $\Gamma_8$ as mentioned above. Region II is limited to concentration interval of $10^{17} \lesssim n_e \lesssim 10^{18}$ см$^{-3}$ below LT. Hence, Weyl semimetal phase already exists in region II. This region is characterised with well-defined chirality, i.e. the formation of a separate Fermi surface for each Weyl node. It enables the additional mechanism of momentum relaxation – intervalley scattering or backscattering [23, 24], thereby leading to a deviation of the mobility from the theoretical value for ionised donor scattering (Fig. 4). However, with reduced $n_e$ the corresponding decrease in $\tau_{tr}$ is accompanied by a decrease in $m_e$ (Fig. 3), similar to region I. As a result, $\mu_H$ depends weakly on $n_e$ in Region II. Thereby, trivial gapless phase still plays noticeable role in carrier kinetics in this region.

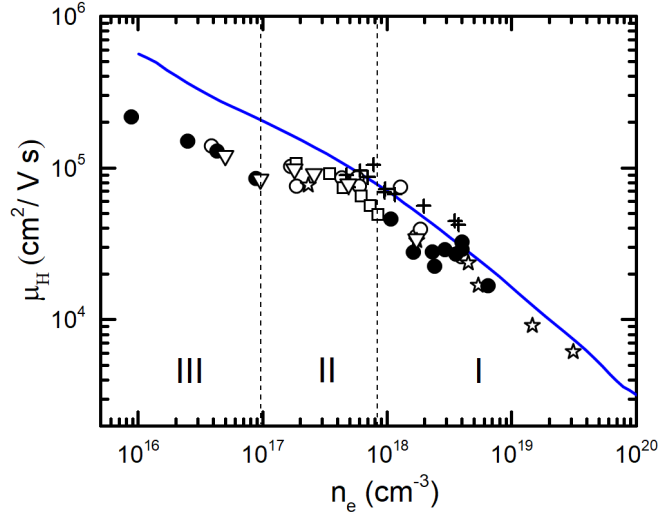

FIG. 4. Electron mobility in HgSe as a function of electron concentration at 4.2K. Filled circles – this work, open circles [9], crosses [17], open squares [11], stars [18], triangles [19]. The solid line is the calculated mobility for single-ionised-donor-scattering mode from [18]. Vertical dashed lines are drawn to separate three regions: Region I – the trivial gapless phase; Region III – the Weyl semimetal phase; Region II – "crossover" (for details see text).

In that sense, this region could be considered as a "crossover" between the Region I and Region III ($n_c < n_e \lesssim 10^{17}$cm$^{-3}$). In Region III, the transport and other properties of HgSe are fully determined by the topological semimetal phase. For example, in all samples from Region III, we observed negative longitudinal MR induced by chiral anomaly and non-trivial manifestation of topological properties in terms of the Hall effect which we will report in future papers. In Region III, $m_e$ does not depend on $n_e$ (Fig. 3). The decrease in $n_e$ with a decrease in concentration of intrinsic donor-type defects in this region should lead to a decrease in the probability of backscattering and therefore to a strong increase in mobility. As seen in Fig. 4 the mobility reaches $2.2 \times 10^5$ cm$^2$/V s at $n_e \approx 9 \times 10^{15}$ cm$^{-3}$, which is a record value for HgSe and is compatible to that of a TaAs when the Fermi level is located near the Weyl node W1 [23]. For the foregoing reasons, the above-mentioned concentration $n_c$ formally corresponds to a boundary, below which the effective mass should tend to zero – as was observed, for example, in graphene when approaching the Dirac point [25]. It is reasonable to assume that reaching $n_c$ in HgSe is unlikely due to the high defect density in this material.

In the conclusion of this subsection, we present the basic electronic parameters obtained from the analysis of the SdH oscillations for HgSe samples from the region III (see Table 1), in which it is most likely to expect the manifestation of the topological properties.

TABLE 1. Parameters of the four investigated samples with the low $n_e$.

| $n_e$ $10^{16}$ cm$^{-3}$ | $m_e/m_0$ | $\tau_{tr}$ $10^{-12}$ s | F T | $A_F$ $10^{-4}$ Å$^{-2}$ | $k_F$ $10^{-3}$ Å$^{-1}$ | $v_F$ $10^5$ m/s | $\varepsilon_F$ meV |
|---|---|---|---|---|---|---|---|
| 0.88 | 0.0326 | 4.01 | 1.25 | 1.19 | 6.16 | 2.19 | 8.9 |
| 2.5 | 0.0330 | 2.50 | 2.60 | 2.48 | 8.88 | 3.12 | 18.2 |
| 4.3 | 0.0345 | 2.54 | 3.85 | 3.68 | 10.8 | 3.63 | 25.8 |
| 8.7 | 0.0334 | 1.63 | 6.19 | 5.90 | 13.7 | 4.75 | 42.9 |

Here $F$ – frequency of SdH oscillations, $n_e$ – electron concentration, $m_e/m_0$ – effective mass ratio, $\tau_{tr} = \mu_H m_e/e$ – transport relaxation time, $A_F = 2\pi e F/\hbar$ – extremal cross-sectional area of the Fermi surface, $k_F = (A_F/\pi)^{\frac{1}{2}}$ – Fermi wave vector, $v_F = (\hbar k_F/m_c)$ – Fermi velocity, and $\varepsilon_F = m_e v_F^2$ – Fermi energy with respect to the Weyl point. The phase of SdH oscillations will be discussed in the separate section for the reasons of its particular importance. As follows from the Table 1, the location of Weyl points is $\approx$ (10–40)meV below the Fermi level within the region III, which is similar to the location of the Weyl nodes in TaAs samples (see Table III in [23]) when the Fermi-level approaches the W1 node.

**B. Oscillation phase**

A distinguishing feature of the Dirac quasiparticle is the nontrivial Berry phase of $\pi$ associated with its cyclotron motion [26, 27]:

$$\Phi_B = \oint_\Gamma \vec{\Omega} \, d\vec{k}, \quad (3)$$

where Γ is the closed electron orbit [the intersection of the constant-energy surface, $\varepsilon(k) =$ const, with the plane, $k_z =$ const, where $z$ is the direction of the magnetic field $\vec{B}$], and $\vec{\Omega}(\vec{k}) = i \int d\vec{k} u_{\vec{k}}^*(\vec{r}) \vec{\nabla}_{\vec{k}} u_{\vec{k}}(\vec{r})$ is the Berry connection; and $u_{\vec{k}}(\vec{r})$ is the amplitude of the Bloch wave function. When the Berry phase is nonzero, the Onsager semiclassical quantization rule [28] becomes [26]:

$$A_F \frac{\hbar}{eB} = 2\pi(n+\gamma) = 2\pi(n + \frac{1}{2} - \frac{\Phi_B}{2\pi}), \quad (4)$$

where $n$ is an integer number. In case of parabolic dispersion law $\Phi_B = 0$, that means phase parameter $\gamma = 1/2$. For an ideal Dirac dispersion $\Phi_B = \pi$ and $\gamma = 0$. A straightforward way to detect the Berry phase in systems with a non-trivial topology is to study SdH oscillations since the basis of this phenomenon consists in a quantisation of orbital motion. Experimentally, a non-trivial $\pi$ Berry phase was obtained from the analysis of SdH oscillations in topological insulators [15, 29-31], Dirac [32, 33] and Weyl [34] semimetals, bulk Rashba semiconductor [35], as well as in graphene [25].

Let us briefly discuss the procedure of extracting the Berry phase from SdH oscillations of transverse MR. We will utilize the following expression for the transverse oscillatory component of MR tensor, adopted from [36] for 3D electrons:

$$\Delta\rho_{xx} \sim -\cos 2\pi\left(\frac{F}{B} + \gamma + \delta\right) = \cos 2\pi\left(\frac{F}{B} + \frac{1}{2} + \gamma + \delta\right), \qquad (5)$$

where $2\pi\delta$ is an additional phase shift with $\delta = 0$ for 2D and $\pm 1/8$ for 3D Fermi surfaces. According to [36], the origin of the minus before cosine follows from the antiphase of the 1st harmonic of magnetic susceptibility with respect to the MR. Also, the minus before cosine can be deduced directly from the first expansion coefficient in the Fourier series of the transverse oscillatory component $\Delta\sigma_{xx}$ of the conductivity tensor [37]. Additionally, it should be taken into account that in a strong magnetic field ($\omega_C \tau_{tr} \gg 1$, where $\omega_C = eB/m_e$ is the cyclotron frequency) $\Delta\sigma_{xx} \sim \Delta\rho_{xx}$ in the lowest order approximation for the small parameter $(\omega_c \tau_{tr})^{-1}$ [37-39]. As it follows from (5), the *n*-th maximum in $\Delta\rho_{xx}$ occurs at $\frac{1}{B_n}$ when:

$$(F/B_n)_{max} = n + 1/2 - (\gamma + \delta) \qquad (6)$$

and the *n*-th minimum in $\Delta\rho_{xx}$ occurs at $\frac{1}{B_n}$ when:

$$(F/B_n)_{min} = n - (\gamma + \delta) \qquad (7)$$

We obtained $\frac{\Phi_B}{2\pi}$ experimentally from the analysis of the Landau-level (LL) fan diagram. The procedure consists in obtaining the intercept of the linear fitting of the $n(\frac{1}{B_n})$ function with *n*-axis, where integer (half-integer) *n* marks minima (maxima) of $\Delta\rho_{xx}(B)$ as it was performed in [25, 32, 40]. In accordance with (6) and (7), this intercept yields the $(\gamma + \delta)$. The values of $\gamma$ between $\frac{3}{8}$ and $\frac{5}{8}$ indicate a trivial Berry phase and values of $\gamma$ between $-\frac{1}{8}$ and $+\frac{1}{8}$ indicate a non-trivial $\pi$-Berry phase. LL fan diagrams for the studied samples are shown in Fig. 5 and the obtained phase shift $\frac{\Phi_B}{2\pi}$ as a function of $n_e$ is shown in Fig. 6. The values of $\frac{\Phi_B}{2\pi}$ for samples with $n_e = 7.1 \times 10^{17}$ and $2 \times 10^{18}$ cm$^{-3}$ were determined from the analysis of SdH oscillations that were reported in [41]. As seen in Fig. 6, there is a pronounced abrupt change in $\frac{\Phi_B}{2\pi}$ from $\approx \frac{1}{2}$ to $\approx 0$ within a narrow concentration interval near $n_e \approx 2 \times 10^{18}$ cm$^{-3}$, indicating the drop in $\Phi_B$ from $\approx \pi$ to $\approx 0$. We consider such behaviour of $\Phi_B$ to be the manifestation of the topological LT in HgSe at critical concentration $n_{LT} \cong 2\times 10^{18}$ cm$^{-3}$.

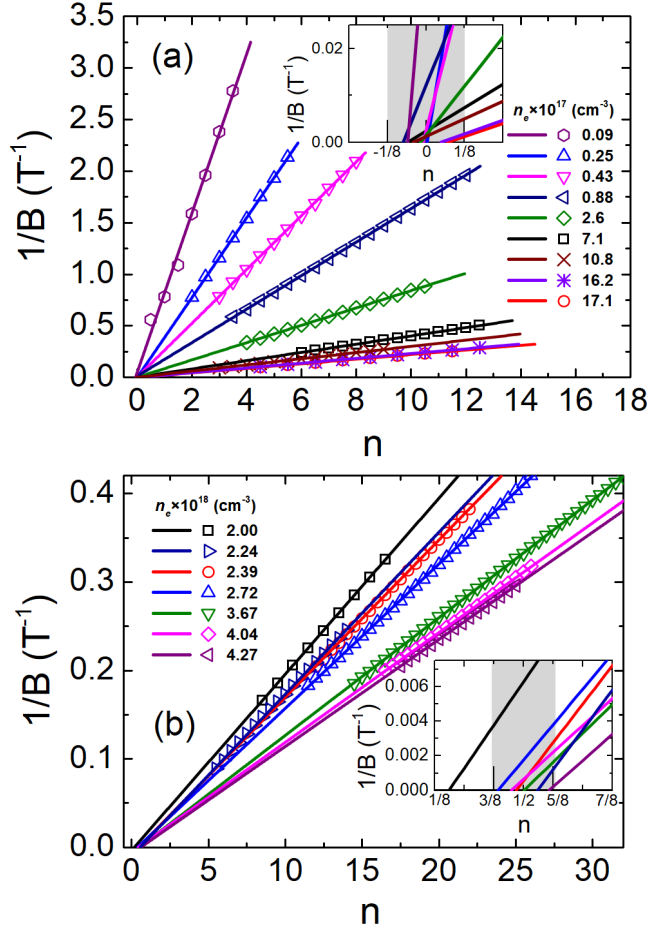

FIG. 5. Landau-level fan diagram for oscillations $\Delta\rho_{xx}(B)$ measured at T=4.2K for HgSe samples with $n_e < n_{LT}$ (a) and $n_e \gtrsim n_{LT}$ (b), where $n_{LT}$ is the critical concentration of the LT. Integers n (half integers n+1/2) are assigned to the minima (maxima) in $\Delta\rho_{xx}$. The insets show a magnified view around the intercept. The intercept γ is either between -1/8 and 1/8 for non- trivial Berry phase [grey area in the inset of panel (a)] or between 3/8 and 5/8 for trivial Berry phase [grey area in the inset of panel (b)]. The only exception is γ ≈ 0.2 for the sample with concentration $n_e = 2\times 10^{18}$ cm$^{-3}$ which we consider to be a critical concentration of the LT.

A similar drop in the phase shift of quantum oscillations near the LT point was recently predicted theoretically in 3D topological semimetals in [39]. It is essential that the flattening of $\mu_H(n_e)$ (Region II in Fig. 4) occurs below approximately the same electron density $n_{LT}$ as was mentioned above. Here it is important to note that, in contradistinction to the TaAs family of WSMs [34], only the W1-type of Weyl cones is supposed in HgSe [4]. Therefore, the phase shift analysis of SdH oscillations in HgSe is greatly simplified due to the absence of additional modes. Thus, the Berry phase may be considered as an appropriate parameter for detection of the LT in WSMs, which occurs by tuning the Fermi level via doping.

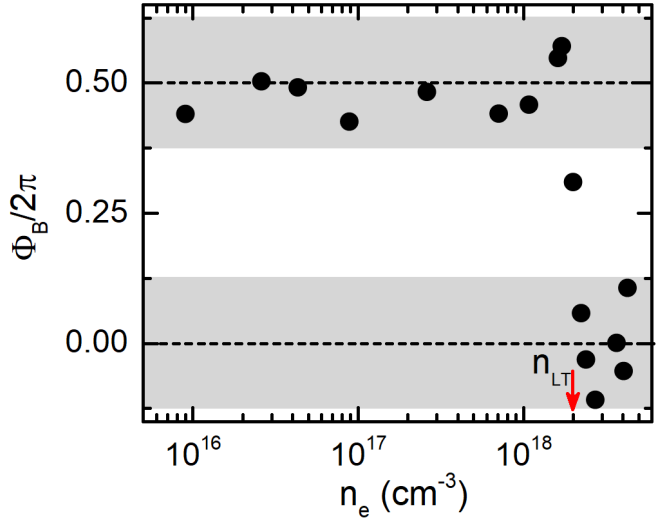

FIG. 6. Phase shift $\frac{\Phi_B}{2\pi}$ of SdH oscillations in HgSe as a function of $n_e$. A pronounced abrupt change in $\frac{\Phi_B}{2\pi}$ from $\approx \frac{1}{2}$ to $\approx 0$ is seen in the vicinity of the concentration $n_{LT} = 2\times10^{18}$ cm$^{-3}$ denoted by the arrow. Upper grey area corresponds to non-trivial Berry phase within δ and the lower one to the trivial Berry phase.

The wave vector $k_{LT}$ for the LT is evaluated as $(2F_{LT}e/\hbar)^{\frac{1}{2}} \approx 0.04$Å$^{-1}$ where $F_{LT}$ is a SdH oscillation frequency for the sample with $n_e = 2\times10^{18}$ cm$^{-3}$. This makes it possible to evaluate the key parameter for HgSe as a WSM candidate – the distance between two Weyl nodes with opposite chirality $\Delta k_W$. Assuming that the separate Fermi spheres of the two Weyl nodes meet at $k_F \approx k_{LT}$, we evaluate $\Delta k_W \cong 2k_{LT} \approx 0.08$Å$^{-1}$. This value is much smaller than the size of the first Brillouin zone in HgSe ($\sim 1$Å$^{-1}$) and is well consistent with $\Delta k_W$ for the TaAs family of WSMs directly measured with ARPES [42].

Furthermore, the type of Weyl nodes is of particular interest. WSMs can be classified into type-I, that exhibit point-like Fermi surface, and type-II with strongly tilted Weyl cones [43]. According to calculations [7], HgTe is in type-II Weyl semimetal phase when the strain is small enough or even zero. However, such calculations were not performed for HgSe. The experimental data presented in current paper also does not provide the opportunity to reach a conclusion on the type of Weyl nodes (type-I or type-II) supposed in unstrained HgSe. In order to elucidate this problem, additional experiments are necessary, for example the measurements of the photocurrent induced by circularly polarized light [44, 45].

## IV. Conclusion

In conclusion, we have demonstrated that studies of the effective mass, electron mobility and phase shift of SdH oscillations in HgSe complement each other. Effective mass as a band spectrum parameter points to a deviation of the dispersion relation $\varepsilon(k)$, given by (2), from an ideal linear dispersion at low energy. In contrast to the case of surface states of topological insulators [14], this deviation seems to be unsubstantial and does not shift the Berry phase from π. The non-trivial Berry phase points to the existence of band touching points in HgSe such as Weyl nodes and related Weyl cones with linear dispersion. The main result is the discovery of the abrupt change ≈ π of the Berry phase (Fig. 6), which we associate with the LT in HgSe that occurs by tuning the Fermi energy via doping. It is relevant argument for a topological semimetal phase in HgSe and is consistent with the results of our previous magnetotransport study [4] of HgSe with low electron concentration – non-centrosymmetric WSM candidate. However, it is obvious that direct experiments, such as ARPES, are necessary to clarify the existence of the Weyl semimetal topological phase in HgSe along with the location and type of Weyl nodes.

## ACKNOWLEDGMENTS

The work was performed as a part of the state task subject "Electron", S.r.№ AAAA-A18-118020190098-5 and the project №18-10-2-6 of the UB RAS Program with support by the RFBR (Project No. 18-32-00198).